\documentclass[preprint]{agujournal2018}
\usepackage{apacite}
\usepackage{url} 
\usepackage{amsmath,amssymb}
\usepackage{mathtools}

\draftfalse

\journalname{Space Weather}

\begin{document}

\title{Spectral Analysis of the September 2017 Solar Energetic Particle Events}

%
%

\authors{A.~Bruno\affil{1}, E.~R.~Christian\affil{1}, G.~A.~de~Nolfo\affil{1}, I.~G.~Richardson\affil{1,2} and J.~M.~Ryan\affil{3}}
\affiliation{1}{Heliophysics Division, NASA Goddard Space Flight Center, Greenbelt, MD, USA.}
\affiliation{2}{Department of Astronomy, University of Maryland, College Park, MD, USA.}
\affiliation{3}{Space Science Center, University of New Hampshire, Durham, NH, USA.}


\correspondingauthor{Alessandro Bruno}{alessandro.bruno-1@nasa.gov}


\vspace{1cm}
\begin{center}
Accepted for publication in \textit{Space Weather}, doi:10.1029/2018SW002085.
\end{center}

%
%

\begin{abstract}
An interval of exceptional solar activity was registered in early September 2017, late in the decay phase of solar cycle 24, involving the complex Active Region 12673 as it rotated across the western hemisphere with respect to Earth. A large number of eruptions occurred between 4--10 September, including four associated with X-class flares. The X9.3 flare on 6 September and the X8.2 flare on 10 September are currently the two largest during cycle 24. Both were accompanied by fast coronal mass ejections and gave rise to solar energetic particle (SEP) events measured by near-Earth spacecraft. In particular, the partially-occulted solar event on 10 September triggered a ground level enhancement (GLE), the second GLE of cycle 24. A further, much less energetic SEP event was recorded on 4 September. In this work we analyze observations by the Advanced Composition Explorer (ACE) and the Geostationary Operational Environmental Satellites (GOES), estimating the SEP event-integrated spectra above 300 keV and carrying out a detailed study of the spectral shape temporal evolution. Derived spectra are characterized by a low-energy break at few/tens of MeV; the 10 September event spectrum, extending up to $\sim$1 GeV, exhibits an additional rollover at several hundred MeV. We discuss the spectral interpretation in the scenario of shock acceleration and in terms of other important external influences related to interplanetary transport and magnetic connectivity, taking advantage of multi-point observations from the Solar Terrestrial Relations Observatory (STEREO). Spectral results are also compared with those obtained for the 17 May 2012 GLE event.
\end{abstract}

%
%

\section{Introduction}\label{Introduction}
It is generally accepted that solar energetic particles (SEPs) are accelerated by a mixture of processes associated with flares and coronal mass ejections (CMEs) (see, e.g. \citet{ref:DESAIGIACALONE2016}).
Such mechanisms are predicted to leave distinct signatures in the energy spectrum, whose measurement thus provides important constraints on SEP origin.
However, spectral features observed at different energies may arise from particle acceleration in different locations (e.g., the flare region, corona or
interplanetary space), so the spectral shapes may exhibit the combined signatures of several dynamic processes that may be complex to disentangle.
Furthermore, the morphology and the evolution of SEP events are strongly influenced by the magnetic connection to sources and 
by interplanetary transport effects and transient/recurrent solar wind (SW) disturbances which significantly complicate the interpretation of spectral measurements. 

The early September 2017 solar events were well-observed by several space- and ground-based instruments, receiving noteworthy attention by a number of papers in the literature (see, e.g., \citet{ref:SUN2017,ref:SHARYKIN2018,ref:GARY2018,ref:OMODEI2018,ref:LONG2018,ref:SEATON2018,ref:WARREN2018,ref:SHEN2018,ref:GUO2018,ref:LUHMANN2018,ref:CHERTOK2018,ref:GOPALSWAMY2018}). In this work we focus on the SEP events that accompany these eruptions, taking advantage of multi-spacecraft data from the Advanced Composition Explorer (ACE) and the Geostationary Operational Environmental Satellites (GOES) to provide an assessment of the SEP spectral shapes over a complete range of energies spanning from few hundreds of keV to a few GeV. 
We also illustrate the effects of SW structures on the SEP spectra.
In addition, observations from the Solar Terrestrial Relations Observatory-Ahead (STEREO-A) are used to provide a more complete view of these SEP events near 1 AU.
The paper is structured as follows: the September 2017 events are introduced in Section \ref{The September 2017 solar events};
in Section \ref{Data} we analyze the various SEP measurements and examine the 
relevant interplanetary data;
Section \ref{SEP spectral analysis} describes the reconstruction and analysis of SEP spectra;
results are presented and discussed in Section \ref{Results};
finally, Section \ref{Summary and conclusions} reports our summary and conclusions.

%
%

\section{The September 2017 solar events}\label{The September 2017 solar events}
The first half of September 2017 was characterized by extreme solar activity mostly related to the complex Active Region (AR) NOAA 12673, which rapidly developed on 4--5 September when near central meridian (e.g., \citet{ref:SUN2017}) and rotated over the west limb on 10 September. A large number of bright eruptions were registered between 4 and 10 September, including 27 associated with M-class flares and four with X-class flares. 
Table \ref{tab:flare_list} lists the $>$M5 flares during this period. That such large AR can emerge late in the declining phase of solar cycles is also demonstrated by the December 2006 events, involving four X-class flares including the powerful X9.0 flare on 5 December and the X3.4 flare on 13 December associated with the 70$^{th}$ ground level enhancement (GLE), linked to AR 10930 during the analogous period of the previous solar cycle \citep{ref:SEP2006}. 

\begin{table}
\centering
\small
\setlength{\tabcolsep}{3pt}
\begin{tabular}{c|c|c|c|c|c|c|c|c|c}
Date & \multicolumn{5}{c|}{Flare} & \multicolumn{4}{c}{CME} \\
& Class & Onset & Peak & End & Location & Speed & 1$^{st}$-app. time & Width & Direction\\
\hline
\textbf{04 Sept.} & \textbf{M5.5} & \textbf{20:28} & \textbf{20:33} & \textbf{20:37} & \textbf{S11W16} & \textbf{1418/1114} & \textbf{20:12/20:36} & \textbf{360/92} & \textbf{S10W10}\\
06 Sept. & X2.2 & 08:57 & 09:10 & 09:17 & S07W33 & 391/260 & 09:48/10:00 & 80/48 & S08W83\\
\textbf{06 Sept.} & \textbf{X9.3} & \textbf{11:53} & \textbf{12:02} & \textbf{12:10} & \textbf{S08W33} & \textbf{1571/1238} & \textbf{12:24/12:24} & \textbf{360/88} & \textbf{S15W23}\\
07 Sept. & M7.3 & 10:11 & 10:15 & 10:18 & S08W47 & 470/597 & 10:24/10:48 & 32/26 & S13W51\\
07 Sept. & X1.3 & 14:20 & 14:36 & 14:55 & S11W49 & 433/477 & 15:12/15:12 & 58/32 & S16W53\\
08 Sept. & M8.1 & 07:40 & 07:49 & 07:58 & S10W57 & 500/450 & 07:36/07:24 & 31/40 & S03W54\\
\textbf{10 Sept.} & \textbf{X8.2} & \textbf{15:35} & \textbf{16:06} & \textbf{16:31} & \textbf{S08W88} & \textbf{3163/2650} & \textbf{16:00/16:09} & \textbf{360/108} & \textbf{S12W85}\\
\hline
\end{tabular}
\caption{List of eruptions associated with major flares ($>$M5.0) originated from AR NOAA 12673 during September 2017. Data in bold refer to the three SEP events registered at Earth. For each event, the flare class, onset/peak/end times (UT) and location (deg) are shown, based on the GOES-15 X-ray archive (\url{ftp://ftp.ngdc.noaa.gov/STP/space-weather/solar-data/solar-features/solar-flares/x-rays/goes/}), along with first appearance time (UT), speed (km s$^{-1}$), angular width (deg) and direction (deg) of the linked CME. The first and the second values reported for CMEs are from the CDAW (\url{https://cdaw.gsfc.nasa.gov/CME_list/}) and the DONKI (\url{https://kauai.ccmc.gsfc.nasa.gov/DONKI/}) catalogs, respectively; CME directions are based on the latter. Sky-plane (space) speeds are reported in case of CDAW (DONKI).}
\label{tab:flare_list}
\end{table}

Three of the major flares, indicated by bold type in Table \ref{tab:flare_list}, were associated with fast CMEs and gave rise to SEP events.
A first, small SEP event observed late on 4 September originated from the moderately intense flare (M5.5) and the geo-effective, halo CME that erupted on the same day. The coordinated data analysis workshops (CDAW, \url{https://cdaw.gsfc.nasa.gov/CME_list/}) catalog of the Large Angle and Spectrometric Coronagraph (LASCO) on board the Solar and Heliospheric Observatory (SOHO) indicates a linear speed of 1418 km s$^{-1}$;
the Database Of Notifications, Knowledge, Information (DONKI, \url{https://kauai.ccmc.gsfc.nasa.gov/DONKI/}) reports a space speed of 1114 km s$^{-1}$ and direction of S10W10, based on the observations of the Sun Earth Connection Coronal and Heliospheric Investigation (SECCHI) instrument on board 
STEREO-A and of SO\-HO/LASCO. Discrepancies in the CME speeds/widths between catalogs are attributable to the different methods used to estimate them including whether they are sky-plane (projected) or space (3-D) speeds based on single- 
or multiple-point coronagraph observations, and the helioradial distances at which they are calculated (see \citet{ref:RICHARDSON2015} and references therein).

The subsequent SEP event was linked to the X9.3 flare peaking at 12:02 UT on 6 September, the largest soft X-ray flare in more than 10 years (since December 2006) and the most intense in cycle 24. It generated strong white-light emission and multiple helioseismic waves observed by the Helioseismic and Magnetic Imager (HMI) on board the Solar Dynamics Observatory (SDO) \citep{ref:SHARYKIN2018}. The explosion was associated with an Earth-directed, nearly symmetrical halo CME with an estimated sky-plane velocity of 1571 km s$^{-1}$ according to the CDAW catalog;
DONKI indicates a 1238 km s$^{-1}$ space speed and a S15W23 direction.
It was also accompanied by an intense and complex radio emission with interplanetary Type II, III and IV bursts, and by long-duration $\gamma$-ray emission.

Finally, a third large SEP event originated following another exceptional flare (X8.2) occurring on 10 September and peaking at 16:06 UT, when the AR NOAA 12673 had just rotated over the western solar limb, so the X-ray intensity may be underestimated due to partial occultation by the limb. To date, it is the second largest soft X-ray flare of cycle 24, and was associated with a very fast (3136 km s$^{-1}$ linear speed) asymmetric halo CME in the CDAW catalog; 
DONKI indicates a space speed of 2650 km s$^{-1}$ and direction of S12W85.
The eruption was accompanied by long-duration emissions at different frequencies, ranging from radio waves (Type II, III and IV bursts) to $\gamma$-rays \citep{ref:GARY2018,ref:OMODEI2018}. Spectacular post-flare coronal loops were observed for nearly a full day. Furthermore, the Solar Ultraviolet Imager (SUVI) on GOES-16 showed evidence of an apparent current sheet associated with magnetic reconnection at the beginning of the eruption, and of an extreme-ultraviolet wave at some of the largest heights ever reported \citep{ref:LONG2018,ref:SEATON2018,ref:WARREN2018}. The resulting SEP event was energetic enough to give rise to a secondary particle shower in the Earth's atmosphere which was subsequently detected by neutron monitors (NMs) on ground as a GLE, the second of solar cycle 24 and the 72$^{nd}$ since NM measurements started in the 1940s (\url{https://gle.oulu.fi/}).

%
%

\section{Data}\label{Data}

\begin{figure}[!t]
\centering
\includegraphics[width=0.95\textwidth]{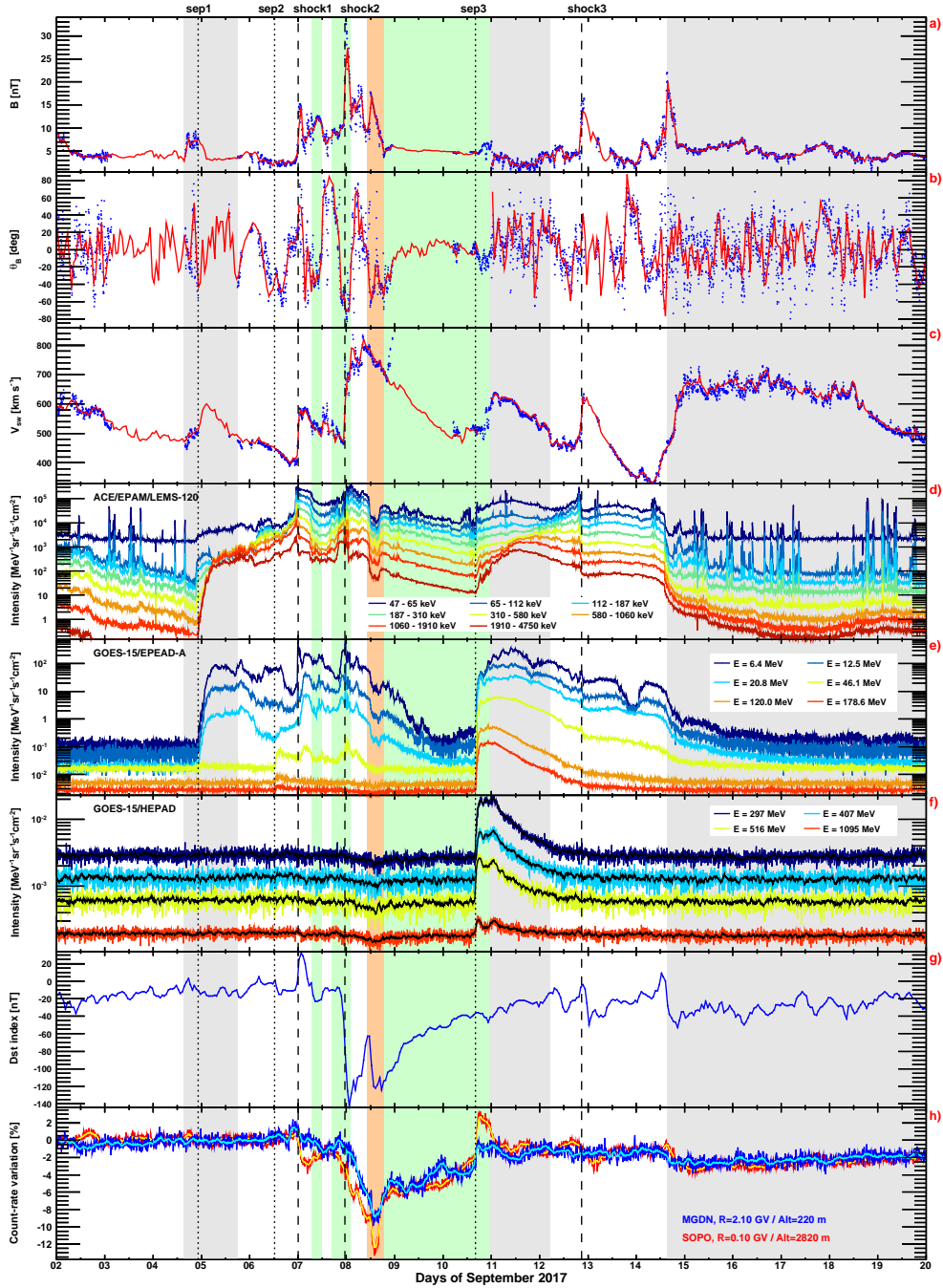}
\caption{From top to bottom: time profiles of IMF intensity (a), IMF latitude (b), SW speed (c), proton intensities measured by ACE/EPAM (d), GOES/EPEAD (e) and GOES/HEPAD (f), $Dst$ index (g), count rate variations registered by SOPO and MGDN NM stations (h). 
Combined ACE and Wind data (red, 1-hr resolution) are superimposed on DSCOVR points (blue, 5-min resolution) in top three panels.
The vertical dotted and dashed lines mark the onset of the SEP events and the time of the shocks, respectively. The green, orange and gray areas indicate the periods of the ICMEs, MC and HSSs, respectively. See the text for details.}
\label{fig:cSW_plus_GEO_plus_SEPs_plus_NM2}
\end{figure}

\subsection{SEP data}\label{SEP data}
\subsubsection{Spacecraft observations}\label{Near-Earth observations}

Figure \ref{fig:cSW_plus_GEO_plus_SEPs_plus_NM2} shows the temporal variation of the relevant interplanetary, geomagnetic and particle data between 2--19 September 2017.
In particular, panels d), e) and f) display the 5-min resolution proton intensities measured by near-Earth spacecraft. Specifically, panel d) reports the observations by the Low Energy Magnetic Spectrometer-120 (LEMS-120) of the Electron, Proton, and Alpha Monitor (EPAM) on board ACE, for 7 energy channels ranging from 47 keV to 4.75 MeV (\url{http://www.srl.caltech.edu/ACE/}). 
Panel e) shows the data from the westward-viewing Energetic Proton, Electron, and Alpha Detector (EPEAD) on board GOES-15; six energy channels (P2--P7) spanning the nominal range 4.2--900 MeV are included. 
Finally, panel f) displays the intensities measured by the four energy channels (P8--P11) of the High Energy Proton and Alpha Detector (HEPAD) on board GOES-15, with a 330--1500 MeV nominal energy interval; the black points correspond to the 1-hr running averages. In case of GOES (\url{https://www.ngdc.noaa.gov/stp/satellite/goes/}), reported mean energy values are based on the calibration schemes by \citet{ref:SANDBERG2014} and \cite{ref:BRUNO_GOES}, respectively below and above 80 MeV.

Vertical dotted lines indicate the onset times of the three SEP events introduced in the previous section, based on a visual inspection of the intensity profile of the GOES highest-energy channel detecting the SEP arrival.
The first enhancement in the proton intensities, registered around 22:00 UT on 4 September and limited to energies below $\sim$150 MeV, originated from the M5.5 flare and the associated full halo CME reported by SOHO/LASCO at 20:12 UT (see Table \ref{tab:flare_list}).
A new increase in the intensities of protons with energies up to a few hundreds of MeV was observed around 12:25 UT on 6 September, related to the X9.3 flare and the linked CME registered by SOHO/LASCO at 12:24 UT.
The temporal evolution of the SEP event is complex and related to interplanetary structures described in Section \ref{IMF, SW and geomagnetic data}.

A third, large SEP event was produced by the X8.2 flare and the associated very fast CME erupting on 10 September, with an onset around 16:05 UT, during the decaying phase of a Forbush decrease (FD). It was energetic enough to give rise to a GLE detected by high-latitude NM stations (see Section \ref{NM data}). 
The sharp increase in proton intensities is consistent with early connection to a shock following a western hemisphere event \citep{ref:CANE1988}, though the W88 location of the event and W85 DONKI CME direction suggest that connection may have been to the eastern flank of the shock assuming nominal Parker spiral interplanetary magnetic field (IMF) lines. However, as pointed out below, the connectivity to the shock is uncertain because of the potential influence of transient SW structures between the Sun and the Earth.
Interestingly, a second peak can be observed in HEPAD profiles at the beginning of 11 September. 
The origin of this feature will be discussed in Section \ref{IMF, SW and geomagnetic data}.

As a final remark, we note that the EPAM/LEMS-120 low-energy channels ($\lesssim$500 keV) are affected by significant electron contamination, as suggested by the gradual enhancement observed apparently before the SEP event onsets. In addition, a number of approximately hour-long bursts can be noted, attributable to ions propagating upstream from the Earth's bow shock when the magnetic connectivity is favorable (see, e.g., \citet{ref:HAGGERTY2000}).

\subsubsection{Neutron monitor observations}\label{NM data}
Panel h) in Figure \ref{fig:cSW_plus_GEO_plus_SEPs_plus_NM2} shows the relative variation in the count-rates registered by 
the South Pole (SOPO, red points) and the Magadan (MGDN, blue points) NM stations, characterized by different values of geomagnetic cutoff rigidity $R$ and altitude (see the legend; \url{http://www.nmdb.eu/}). 
For SOPO $R$ is negligible and the effective detection threshold is determined by the atmospheric cutoff ($\sim$300 MeV).

The error bars refer to the statistical uncertainties. The yellow/cyan points denote the corresponding 1-hr running averages. 
The SEP event on 10 September gave rise to a GLE, the second of solar cycle 24,
commencing at $\sim$16:10 UT during the decaying phase of a major FD, and lasting for several hours.
It was a relatively small GLE event, as the maximum relative increase in the SOPO count-rates was $\sim$6\%. 
The two-peak structure observed in the HEPAD profiles is also evident in the relatively high-cutoff stations, including MGDN.

\subsection{Interplanetary and geomagnetic data}\label{IMF, SW and geomagnetic data}
The aim of this section is to describe the SW structures influencing the near-Earth environment in early September 2017, and help to interpret the particle
observations discussed in the previous sections.
In particular, the profile of the IMF intensity, the IMF latitude in GSE coordinates and the SW speed are reported in panels a), b) and c) of Figure \ref{fig:cSW_plus_GEO_plus_SEPs_plus_NM2}, respectively. Data are based on the OMNIWeb database (\url{http://OMNIWeb.gsfc.nasa.gov}), which provides in-situ observations time-shifted to the bow shock nose of the Earth \citep{ref:KING_PAPITASHVILI2004}. Specifically, combined ACE and Wind data (red, 1-hr resolution) are superimposed on DSCOVR points (blue, 5-min resolution). 
Gray shading indicates corotating high speed streams (HSSs), while the green regions are interplanetary CMEs (ICMEs; see, e.g., \citet{ref:ZURBUCHEN2006,ref:KILPUA2017} and references therein); as discussed below, the orange shading emphasizes the presence of a magnetic cloud (MD) structure. 

Three interplanetary shocks passed by during this interval at the times indicated by the vertical dashed lines. 
The first shock, marked by the commencement of a minor geomagnetic storm at 23:43 UT on 6 September, as evident in the temporal profile of the $Dst$ index reported in panel g) of Figure \ref{fig:cSW_plus_GEO_plus_SEPs_plus_NM2}, was driven by the interplanetary counterpart of the CME observed by SOHO/LASCO on 4 September at $\sim$19 UT and associated with the first SEP event which shows a local enhancement at low energies in the vicinity of the shock. The first ICME interval indicated (shaded green) following the shock was suggested by \citet{ref:SHEN2018}, though the usual SW temperature ($Tp$) decrease \citep{ref:RICHARDSON1995} was not present, and it was associated with a decrease in the low-energy particle intensity enhancement associated with this shock. The second ICME interval, following this shock and commencing at $\sim$19:40 UT, did have a clear $Tp$ relative reduction (and increase in the helium-proton ratio) and was present at Earth at the time of arrival of the second shock, at 23:00 UT on 7 September (based on the storm sudden commencement time). This shock was associated with the CME 
observed by SOHO/LASCO on 6 September at 12:24 UT that was also associated with the second SEP event in Figure \ref{fig:cSW_plus_GEO_plus_SEPs_plus_NM2}. Again there is a low-energy particle enhancement in the vicinity of this shock.
An intense geomagnetic storm occurred with $Dst$ reaching -124 nT early on 8 September, as displayed in panel g) of Figure \ref{fig:cSW_plus_GEO_plus_SEPs_plus_NM2},
following strong ($\sim$30 nT) southward (negative latitude, see panel b) magnetic fields that were caused by the second shock compressing the southward fields in the ICME through which it was propagating. 

The ICME following this shock had two components. The first, marked by the orange shading in Figure \ref{fig:cSW_plus_GEO_plus_SEPs_plus_NM2}, exhibited many of the signatures of a magnetic cloud (MC) (e.g., \citet{ref:KLEIN1982}), including a distinct enhanced but declining IMF intensity, declining SW speed, and low $Tp$, as well as enhanced He/proton ratio and oxygen charge states, 
and bi-directional suprathermal electron beams. However, there was no significant rotation of the IMF vector, so it may be termed a ``MC-like'' ICME \citep{ref:WU2015}; for brevity, we will refer to this region as the ``MC'' (shaded orange). It was followed by a second, extended ICME structure (green shading) characterized by a low variance, slightly enhanced, near-radial sunward magnetic field, depressed $Tp$, a continuing decline in SW speed, and bidirectional suprathermal electrons. 
Following a recovery as the field turned temporarily northward, a second peak in $Dst$ (-109 nT) was driven by southward fields ($\sim$17 nT) inside the MC. Then, a recovery occurred as the field returned northward in the following region of this ICME (shaded green). 
There is a gap in the OMNIWeb data near the end of this region, but the DSCOVR data suggest that it extended to $\sim$00 UT on 11 September based on the end of this region of low variance, near-radial, magnetic field. 
This ICME was followed by a brief HSS (gray shading on 11--12 September) probably attributed to a weak influence from a negative polarity coronal hole. 
The SEP data show a local decrease during passage of the MC at all energies from tens of keV to the peak of the FD observed by NMs. 

A third shock on 12 September at $\sim$20:02 UT (storm commencement time) was likely produced by the passage of the eastern flank of the shock associated with the 10 September event. This is consistent with the glancing blow with an arrival time of 13 September, $\sim$02 UT$\pm$7 hours based in ENLIL+CONE modeling indicated in the DONKI database. However, closer examination of the SW data indicates that this was not a fully-steepened shock. The subsequent lack of ICME-like signatures, in particular low $Tp$, indicates that the associated ICME did not encounter Earth, consistent with the far western origin of this event. 
Finally, a long-duration HSS was observed on 14 September, 
probably associated with the low-latitude extension of the northern polar coronal hole that passed central meridian on 10 September. It carried an intermittent southward IMF and its effect on the Earth endured for several days, triggering a moderate
geomagnetic storm.
The SEP data show an enhancement at the lowest energies in the vicinity of the shock, and also a rapid intensity decrease with the arrival of the HSS on September 14 which terminated the event at low energies (below few hundreds of keV), while an extended decay, already started before the HSS passage, can be observed at higher energies.

Returning to the onset of the 10 September event, this evidently occurred close to the time when Earth was moving from an ICME to a HSS, so we suggest that the double peak in the particle intensity at the highest energies may be associated with this transition, resulting in an improved connection to the particle source.
This feature is less evident at lower energies. Possible reasons may be that the source of the high-energy particles was more spatially confined, and hence connectivity was more critical for the detection of particles, and the low-energy particle intensities were still rising when Earth exited the ICME whereas the highest energies had started to decay. \citet{ref:GUO2018} also proposed a second particle injection at the shock through merging of the ICME associated with the 10 September event with the two ICMEs that originated on 9 September from the same AR with similar directions. However, there does not appear to be evidence of a second particle injection in the available radio data from STEREO-A or Wind, that clearly show only emissions associated with the original onset of the SEP event.

\begin{figure}[!t]
\centering
\includegraphics[width=\textwidth]{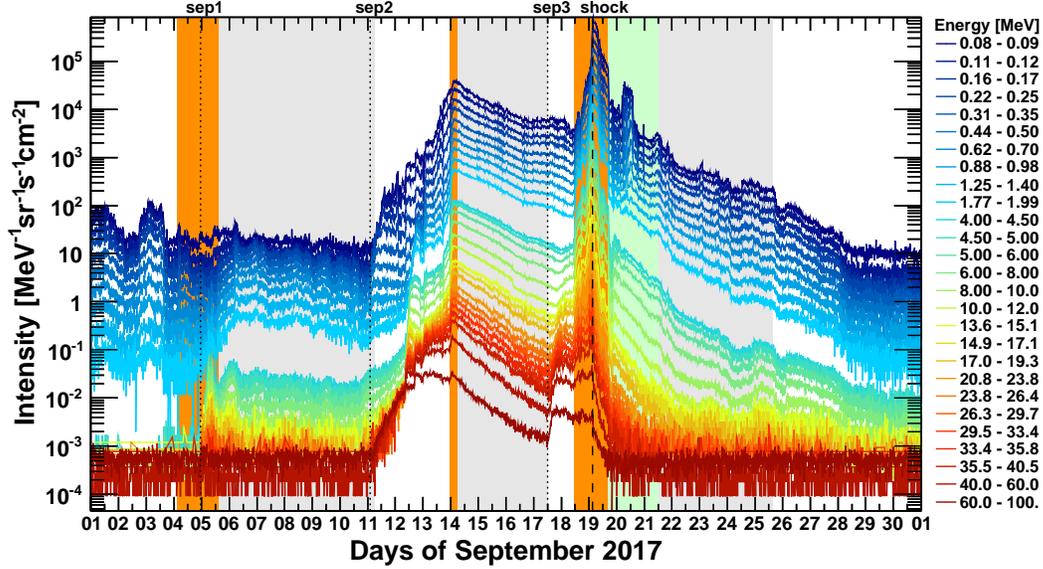}
\caption{Temporal profiles of proton intensities measured by the SEPT, LET and HET instruments on board STEREO-A during September 2017. The vertical dotted and dashed lines mark the onset of the SEP events and the time of the shock, respectively. The green and gray areas indicate the periods of the ICMEs and HSSs, respectively. In this case, the orange shading marks the CIRs.}
\label{fig:stereo}
\end{figure}
\begin{figure}[!t]
\centering
\includegraphics[width=0.5\textwidth]{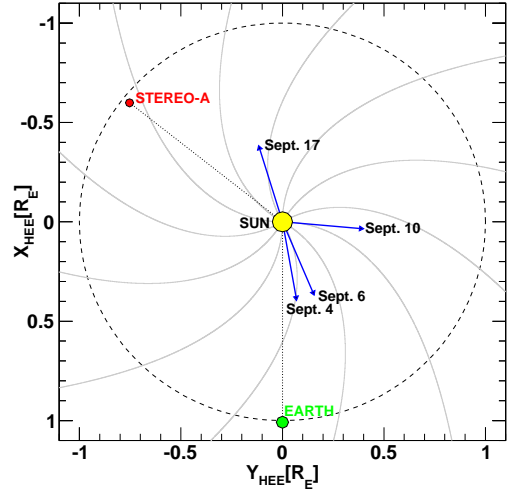}
\caption{Location of the Earth (ACE, GOES) and STEREO-A in Heliocentric Earth Ecliptic (HEE) coordinates during September 2017. The arrows indicate the direction of the CMEs associated with the SEP events observed at Earth/STEREO-A. The nominal Parker-spiral IMF lines assuming $V_{SW}$=450 km s$^{-1}$ are also reported.}
\label{fig:spacecraft_positions}
\end{figure}

\subsection{Stereo observations}\label{Stereo observations}
STEREO-A observations during this period made $\sim$128 deg east of Earth (see Figure \ref{fig:spacecraft_positions}) provide additional information on the SEP events discussed above and their longitudinal extent.
Figure \ref{fig:stereo} displays the temporal profiles of proton intensities measured 
by the Solar Electron and Proton Telescope (SEPT; 0.084--6.5 MeV, 10-min resolution), the Low Energy Telescope (LET; 4--12 MeV, 10-min resolution) and the High Energy Telescope (HET; 13.6--100 MeV, 15-min resolution). In case of SEPT, only selected channels are shown for the sake of simplicity. 
As in Figure \ref{fig:cSW_plus_GEO_plus_SEPs_plus_NM2}, the grey shading indicates HSSs observed at STEREO-A, but here, orange shading indicates corotating interaction regions (CIRs) at the stream leading edges, inferred from inspection of the STEREO-A plasma and magnetic field data, not shown here.

The initial SEP enhancement in Figure \ref{fig:stereo} was associated with the 4 September event, at $\sim$W143 deg relative to the spacecraft longitude, while it was passing through the CIR marking the arrival of a HSS. The prompt rise in the proton intensity suggests that particles propagated rapidly from the eastern flank of the shock.
There is a hint of an increase from the 6 September event, but it is not compelling on the ongoing event. A significant enhancement was registered early on 11 September, demonstrating that the 10 September event was very broad in longitude even at high energies, as the parent flare was located at $\sim$E145 deg relative to STEREO-A. In this case the magnetic footpoints of STEREO-A were connected to the western flank of the shock, and measured intensities exhibit a much more gradual increase. The delayed arrival ($>$10 hours later than the flare onset) may be attributed to cross-field diffusion in the SW. The event duration can be inferred to be much longer with respect to near-Earth observations, well beyond the onset of another high-energy event occurring on 17 September at $\sim$12 UT from the same AR when at $\sim$W167 ($\sim$E40 of STEREO-A), that evidently was not observed at Earth. This event was linked to a fast halo CME with a 1385 (1404) km s$^{-1}$ speed according to the CDAW (DONKI) catalog. 

An interesting feature is the non-energy-dispersive increase in intensity early on 14 September which was associated with entry into -- crossing of the stream interface -- a corotating HSS. This suggests that connection to the particle event and/or particle transport in longitude was more favorable in the stream than in the preceding SW. 
In particular, a study based on the solar energetic particle event modeling (SEPMOD) of this event \citep{ref:LUHMANN2018} suggests that STEREO-A may have become connected to the shock associated with the 10 September event beyond 1 AU at this time. Thus, the observations suggest that field lines in the HSS were connected to this shock, but those in the preceding slow SW were more poorly connected.

An interplanetary shock arrived on 19 September at 02:56 UT, when STEREO-A was passing a CIR. At the same, the SW speed exceeded 800 km s$^{-1}$ and a significant enhancement of low-energy protons was observed. The CIR was followed by the arrival of an ICME, as suggested by the drop in density and temperature, and an enhanced field with a rotation, followed by a weaker, smoother field. The ICME caused a FD of proton intensities. Then another HSS reached the spacecraft. Such interpretation is supported by the results of the ENLIL\-+\-CONE model in DONKI, with the flank of the ICME passing STEREO-A at the time of a stream leading edge.

%
%

\section{SEP spectral analysis}\label{SEP spectral analysis}
In this section, the SEP observations introduced above will be used to construct energy spectra over a wide energy range.
The GOES data are affected by significant uncertainties related to the poor resolution of the detector and high contamination by out-of-acceptance particles \citep{ref:BRUNO_GOES}. 
In addition, the intensities measured by the HEPAD channels and, to a lesser extent, the highest energy channels of the EPEADs, include 
a high background associated with galactic cosmic rays (GCRs). 

To improve the reliability of the EPEAD/HEPAD spectroscopic measurements, we take advantage of two different cross-calibration schemes. For the data points below 80 MeV (P2--P5 channels), the mean energies by \citet{ref:SANDBERG2014} are used, 
based on a calibration study of the Energetic Particle Sensors (EPSs) on board GOES-5, -7, -8, and -11, using as reference the observations of the Goddard Medium Energy (GME) experiment on board the Interplanetary Monitoring Platform-8 (IMP-8); the derived cross-calibrated energies have been validated by \citet{ref:RODRIGUEZ2017} by comparison with the STEREO data. 
A background correction is applied by subtracting the minimum intensity measured during the 30-day interval prior to the SEP events, based on 6-hr moving averaged data; conservatively, a 20\% systematic uncertainty is assumed. 
To avoid east-west effects \citep{ref:RODRIGUEZ2010}, more relevant at lower energies, 
only observations from the westward viewing EPEADs are used.

The GOES data points above 80 MeV are based on \citet{ref:BRUNO_GOES}, who took advantage of the SEP measurements of the Payload for Antimatter Matter Exploration and Light-nuclei Astrophysics (PAMELA) \citep{ref:BRUNO2018} to calibrate the two most energetic channels (P6--P7) of the EPEADs and the four HEPAD channels (P8--P11), for both GOES-13 and -15 units. 
As east-west effects are negligible at high energies, data from both westward and eastward looking EPEADs are used in this range. 
A background correction is applied by subtracting the average intensity measured during the 24-hr quiet solar period prior to the SEP events. 
It should be noted that derived ``effective'' mean energies represent average values and do not account for spectral index variations.
A 20\% (30\%) systematic uncertainty is assumed for the EPEAD (HEPAD) points, based on the comparison with PAMELA measurements \citep{ref:BRUNO_GOES}.

In case of ACE and STEREO instruments, the background in each energy bin is evaluated as the minimum intensity measured during a 30-day interval prior to the SEP events, based on 6-hr moving averaged data. To a first approximation, the mean energy values are obtained by estimating the logarithmic center of each bin. 
However, since the two highest-energy channels of HET span a relatively much wider range (40--60 MeV and 60--100 MeV, respectively),
the corresponding ``true'' mean energies are significantly affected by spectral shape variations and, thus, the above assumption is no longer reasonable.
Consequently, a different approach based on \citet{ref:LAFFERTY1995} is used in this case:
\begin{equation}\label{eq:BAND}
E_{mean} = \left[ \frac{E_{max}^{1-\gamma}-E_{min}^{1-\gamma}}{(E_{max}-E_{min})(1-\gamma)} \right]^{-\frac{1}{\gamma}},
\end{equation}
where $E_{min}$ and $E_{max}$ are the channel lower and upper energy limits,
and $\gamma$ is the spectral index derived by the power-law fit of HET spectral points between 30--40 MeV.

The ``spikes'' in the ACE temporal profiles of intensities, attributable to ions propagating upstream from the Earth's bow shock (see Section \ref{Near-Earth observations}), are removed. Since the lowest energy channels are affected by electron contamination, only the intensities above 300 keV are considered; in addition, a 20\% systematic uncertainty is associated with the data points.

In general, statistical errors are evaluated by accounting for the GCR background subtraction, by using 68.27\% confidence level intervals for Poisson signal/background distributions according to \citet{ref:FELDMANCOUSINS1998}. Statistical and systematic uncertainties are summed in quadrature.

Event-integrated energy spectra are obtained by summing up the SEP intensities measured in each energy bin over the event duration. The integration interval is computed by identifying the event start/stop times in the intensity temporal profiles. When a new event commences while a preceding one was still in progress, the onset time of the second event is set as the end time of the first event. Consequently, the spectrum for the second event will include a contribution from the decay of the previous event. Finally, it should be noted that, since the background correction is based on pre-event intensities, SEP event-integrated intensities are somewhat underestimated -- especially above several tens of MeV -- if FD periods are present, such as during the decaying phase of the 6 September event and the initial phase of the 10 September event. 

\subsection{Spectral fits}\label{Spectral fits}
In order to characterize the estimated event-integrated energy spectra, we fit them with several spectral shapes. 
A first, purely empirical model is given by the double power-law function
by \citet{ref:BAND1993} (hereafter Band function):
\begin{equation}\label{eq:BAND}
\Phi_{Band}(E) = 
\begin{dcases*} 
A \hspace{0.1cm} E^{-\gamma_{a}} \hspace{0.1cm} exp\left(-E/E_{0}\right) \hspace{3.98cm} \text{for $E<(\gamma_{b}-\gamma_{a}) \hspace{0.1cm} E_{0}$},\\
A \hspace{0.1cm} E^{-\gamma_{b}} \hspace{0.1cm} \left[(\gamma_{b}-\gamma_{a})\hspace{0.1cm} E_{0}\right]^{(\gamma_{b}-\gamma_{a})} \hspace{0.1cm}exp\left(\gamma_{a}-\gamma_{b}\right) \hspace{0.7cm} \text{for $E>(\gamma_{b}-\gamma_{a})\hspace{0.1cm}E_{0}$},
\end{dcases*}
\end{equation}
originally developed to fit gamma-ray burst spectra. It is defined by four free parameters ($A$, $\gamma_{a}$, $\gamma_{b}$, $E_{0}$), providing a smooth transition between two energy regions characterized by different spectral indices ($\gamma_{a}$ and $\gamma_{b}$); the transition energy is given by $(\gamma_{b}-\gamma_{a}) \hspace{0.1cm} E_{0}$. While such spectral breaks, typically occurring at energies of few tens of MeV, have been often associated with the limits of shock acceleration (see, e.g., \citet{ref:DESAI2016} and references therein), they can be explained by accounting for interplanetary transport effects \citep{ref:LILEE2015,ref:ZHAO2016}.

A second functional form is based on \citet{ref:ELLISON_RAMATY1985} (hereafter referred as E-R), and consists of a power-law spectrum modulated by an exponential: 
\begin{equation}\label{eq:E-R_function}
\Phi_{E-R}(E) = A \hspace{0.1cm} E^{-\gamma} \hspace{0.1cm} exp\left(-E/E_{r}\right),
\end{equation}
where $E_{r}$ is the cutoff or rollover energy. In the scenario of diffusive shock acceleration, the spectral rollover is attributed to particles escaping the shock region during acceleration due to effects mostly related to the limited extension and lifetime of the shock \citep{ref:LEERYAN86,ref:LEE2005}. This function has been recently used by \citet{ref:BRUNO2018} to fit the time-integrated energy spectra of the high-energy ($>$80 MeV) SEP events observed by the PAMELA experiment.

In general, multiple spectral features can be present at different energies, and the above functional forms hardly reproduce the spectral shapes over the complete energy range of SEPs. In particular, the Band function reasonably describes the SEP spectra below several tens of MeV, but it reduces to a single power-law extending to infinity for energies much larger than the break energy; consequently, it can not be used to account for the high-energy (hundreds of MeV) spectral rollovers recently found in PAMELA observations \citep{ref:BRUNO2018}. 
In order to reproduce both the low-energy break and the high-energy rollover in the SEP spectra, Equations \ref{eq:BAND} and \ref{eq:E-R_function} can be combined into:
\begin{equation}\label{eq:BANDER}
\Phi_{tot}(E) = \Phi_{Band}(E) \hspace{0.1cm} exp\left(-E/E_{r}\right),
\end{equation}
i.e. a double-power law (Band) function multiplied by an (E-R) exponential cutoff.
Hereafter we refer to the above functional form as the ``combined'' function.

As a final remark we note that, overall, significant cross-correlations may exist between the fit parameters, in particular between the break/rollover energies and the spectral indices \citep{ref:DESAI2016,ref:BRUNO2018}, resulting in large parameter uncertainties.
Fit errors are evaluated with the MINOS technique (see, e.g., \citet{ref:FERBEL1983}).

\begin{figure}[!t]
\centering
\includegraphics[width=0.73\textwidth]{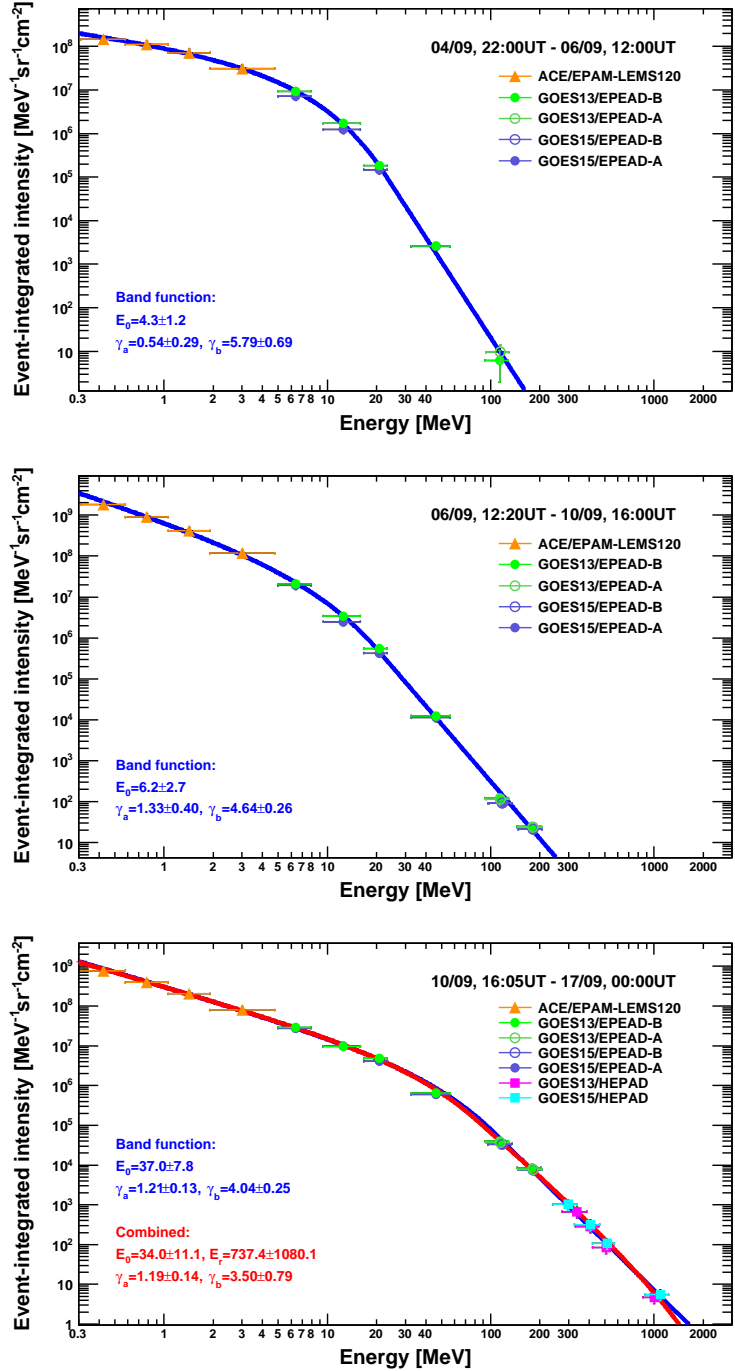} 
\caption{The time-integrated energy spectra of the 4, 6 and 10 September 2017 SEP events (top, middle and bottom panel, respectively) measured by ACE and GOES-13/15. The vertical error bars account for statistical and systematic uncertainties. 
The horizontal error bars denote the channel nominal/effective energy ranges.
The blue and the red curves denote the fits performed by using the Band and the combined functions. The integration intervals, along with fit parameters and associated uncertainties are also reported with the same color code.}
\label{fig:Fluences}
\end{figure}
%
%

\section{Results}\label{Results}
The time-integrated energy spectra of the 4 and 6 September 2017 SEP events measured by ACE and GOES-13/15 above 300 keV are shown in top and middle panels of Figure \ref{fig:Fluences}, respectively. 
The vertical error bars account for both statistical and systematic uncertainties.
The horizontal error bars denote the nominal energy ranges or, in the case of GOES, the ``effective'' energy ranges 
estimated by \citet{ref:SANDBERG2014} and \cite{ref:BRUNO_GOES}. The curves indicate the fits performed with the Band function; the fit parameters along with associated uncertainties are also reported. The Band function provides good fits to the spectra, which are very soft ($\gamma_{b}$$\approx$5.8 and $\gamma_{b}$$\approx$4.6, respectively) above the break energy (4.3 MeV and 6.2 MeV, respectively). In addition, the 4 September spectrum is almost flat below the break ($\gamma_{a}$$\approx$0.5). As reconstructed spectra are limited to energies below $\sim$150 MeV and $\sim$200 MeV, respectively, no reliable assumption can be made regarding an high-energy spectral rollover.

In contrast, as demonstrated in the bottom panel of Figure \ref{fig:Fluences}, the spectrum measured for the 10 September SEP event extends up to $\sim$1 GeV. Since faster shocks can accelerate particles to higher energies, 
the high energies reached in the 10 September event are consistent with the associated ultra-fast CME (see Table \ref{tab:flare_list}). 
In addition, in comparison to 4 and 6 September events,
a powerful radio emission at higher frequencies accompanied the event \citep{ref:CHERTOK2018}, implying that SEPs were accelerated closer to the Sun, where the magnetic field is more intense and hence the maximum energy to which SEPs can be accelerated is higher \citep{ref:ZANK2000,ref:GOPALSWAMY2017}. 
\citet{ref:GOPALSWAMY2018} estimated a shock height of 1.4 Rs at Type II onset, in agreement with previous GLE observations.
For comparison, the steeper radio spectrum with a peak at lower frequencies measured during the 4 September event is indicative of a post-eruption origin, while the 6 September event had intermediate features \citep{ref:CHERTOK2018}.

\begin{figure}[!t]
\centering
\includegraphics[width=\textwidth]{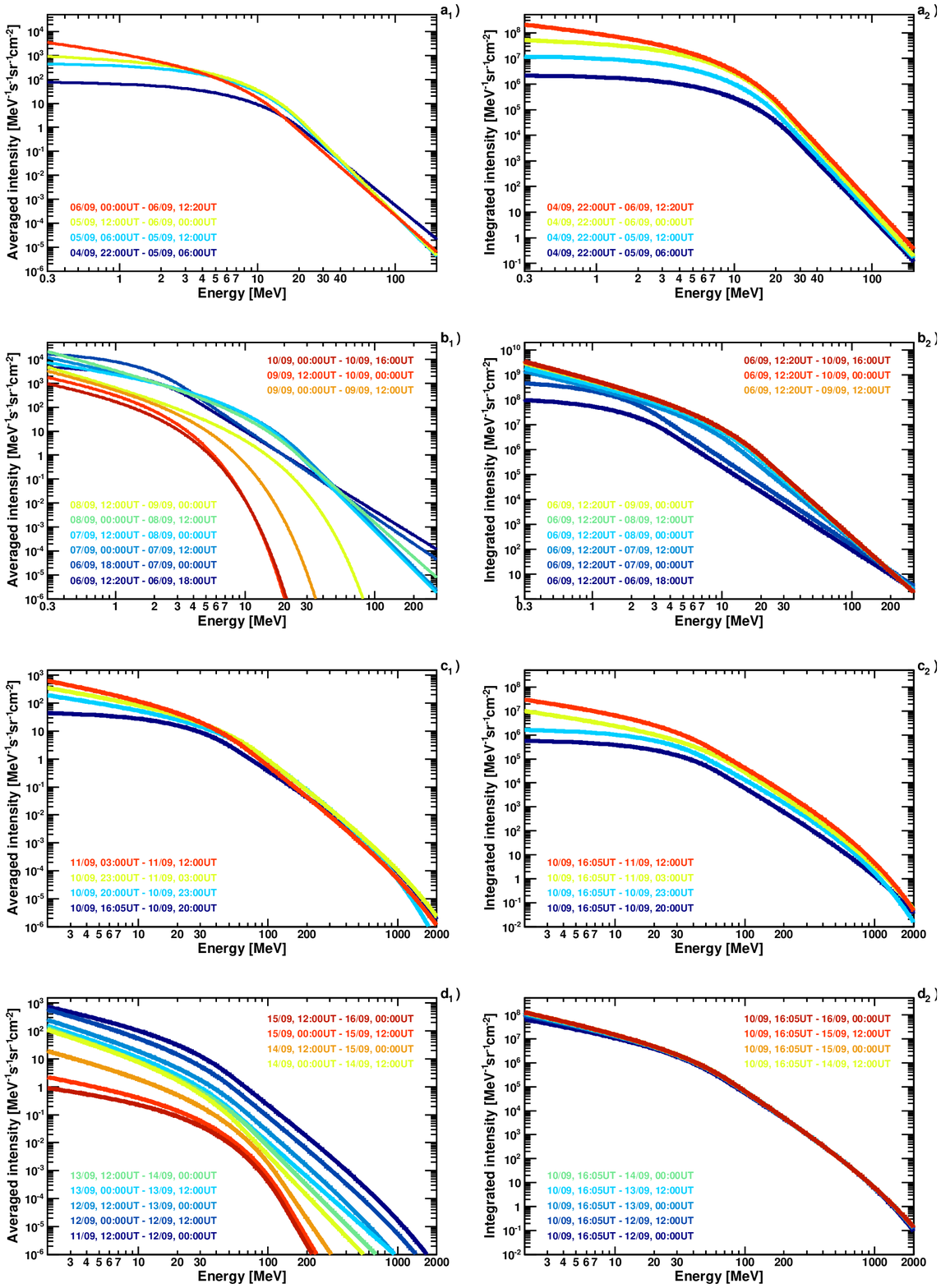} 
\caption{Spectral fits obtained for the 4 September event (a), the 6 September event (b) and the long-duration 10 September event (c and d). 
Left panels are based on the energy spectra averaged during successive time intervals, while right panels show the fits of the corresponding spectra integrated over cumulative intervals, with same color code (see labels).}
\label{fig:spectra_evolution}
\end{figure}

The high-energy data in the spectrum of the 10 September event suggest the presence of a rollover -- albeit with large uncertainties due to the few points and their error bars -- similar to that found in the high-energy SEP observations reported by the PAMELA mission \citep{ref:BRUNO2018}, that may be consistent with the limits of diffusive shock acceleration (see Section \ref{Spectral fits}). 
Comparing the fits performed with the Band (blue) and the combined (red curve) functions, we
obtain a $\sim$1.36 value for the ratio of the corresponding reduced $\chi^{2}$ ($F$-test).
Therefore the spectral shape is better reproduced by the latter functional form, which provides a reasonable fit of the data points in the full energy range accounting for both the low-energy break (34 MeV) and the high-energy rollover (737 MeV).
However, the interpretation of spectra shapes is significantly complicated by a series of overlapping events and related interplanetary structures (local shocks, ICMEs and HSSs), as discussed in Section \ref{IMF, SW and geomagnetic data}, influencing SEP intensities hence spectra. Consequently, it is not realistic to account for the spectral features only in terms of particle acceleration.

\begin{figure}[!t]
\centering
\includegraphics[width=\textwidth]{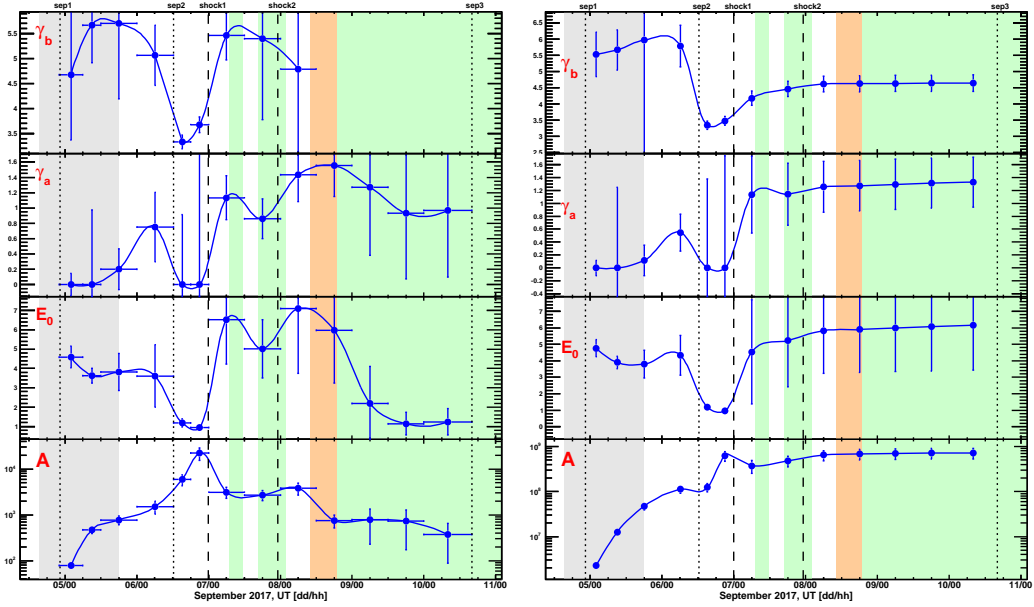} 
\caption{Left - Evolution of the Band fit parameters for the average spectra of the 4 and 6 September 2017 events reported in left panels of Figure \ref{fig:spectra_evolution}. 
Right - Evolution of the Band fit parameters for the cumulative spectra of the 4 and 6 September 2017 events reported in right panels of Figure \ref{fig:spectra_evolution}. 
The curves are to guide the eye. 
The vertical error bars account for fit parameter uncertainties.
The vertical dotted and dashed lines mark the onset of the SEP events and the time of the shocks, respectively. The green, orange and gray areas indicate the periods of the ICMEs, MC and HSSs, respectively.}
\label{fig:cParTime_Sept06}
\end{figure}

\begin{figure}[!t]
\centering
\includegraphics[width=\textwidth]{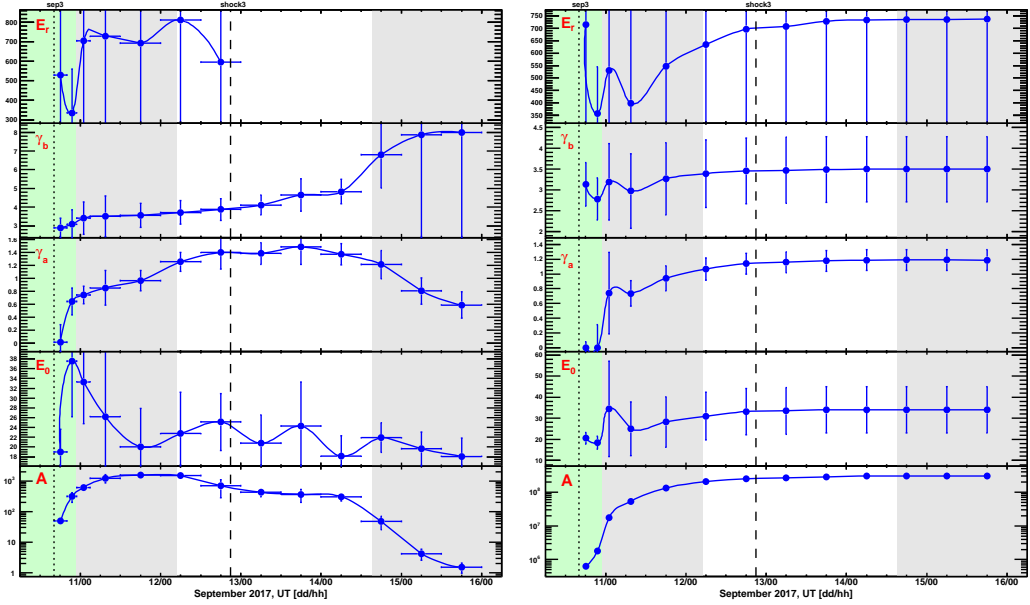} 
\caption{Left - Evolution of the combined fit parameters for the average spectra of the 10 September 2017 event reported in left panels of Figure \ref{fig:spectra_evolution}. 
Right - Evolution of the combined fit parameters for the cumulative spectra of the 10 September 2017 event reported in right panels of Figure \ref{fig:spectra_evolution}. 
The vertical error bars account for fit parameter uncertainties.
The curves are to guide the eye. The vertical dotted and dashed lines mark the onset of the SEP event and the time of the shock, respectively. The green and gray areas indicate the periods of the ICMEs and HSSs, respectively.}
\label{fig:cParTime}
\end{figure}

\subsection{Spectra temporal evolution}
The left panels in Figure \ref{fig:spectra_evolution} display the fits of the SEP spectra obtained in successive time intervals during the 4 September event (a), the 6 September event (b) and the long-duration 10 September event (c and d). The fits for the 4 and 6 September events are based on the Band function, while the combined functional form was used for the 10 September event. The spectra are evaluated by averaging intensities on a 12-hr timescale; a higher time resolution (3--6 hours) is used during the initial phase of the events (see labels). In addition, only data above 2 MeV are included for the 10 September event due to the difficulty in fitting the complete energy spectrum, which exhibits a further softening at lower energies in the early phase attributable to a low energy component from the previous event. The time variations of the fit parameters are summarized in left panels of Figures \ref{fig:cParTime_Sept06} and \ref{fig:cParTime}. It should be stressed that fit parameters are typically correlated. The right-hand panels of Figure \ref{fig:spectra_evolution} show the cumulative spectra for each event integrated up to the end time of each spectrum in the left panels and indicated with the same color code. The corresponding fits to the cumulative spectra are shown in the right panels of Figures \ref{fig:cParTime_Sept06} and \ref{fig:cParTime}. 

The initial phase of the 4 September event -- as well as the other events -- was characterized by velocity dispersion effects, with higher-energy particles arriving earlier, resulting in relatively hard spectra. The spectra was almost flat at low-energies ($\gamma_{a}$$\approx$0). In the subsequent three intervals the high-energy part of the spectrum did not change significantly, in particular the break energy remained constant, while the low-energy spectrum became softer due to the increasing intensities.

The spectral evolution of the 6 September event can be divided into three phases. During the first one (first two time bins), the break energy was very low ($E_{0}$$\approx$1 MeV) and the spectrum was flat ($\gamma_{a}$=0) and relatively hard ($\gamma_{b}$$\approx$3.5) in the energy ranges below and above the spectral transition, respectively. 
Derived spectra, especially at low energies, include a particle component associated with the ongoing 4 September event, along with the related shock. 
The second phase (subsequent three time bins) commenced after the arrival of the interplanetary shock at the end of 6 September: the break energy increased (5--6 MeV) and the spectrum became softer ($\gamma_{a}$$\approx$1 and $\gamma_{b}$$\approx$5). The arrival of the shock-ICME complex structure at the end of 7 September caused large FD effects, inducing an enhancement of $E_{0}$ and $\gamma_{a}$.
The third phase (last four time bins) started with arrival of the MC, corresponding to the peak of the FD, and extended over its decaying phase up the onset of the following SEP event. At the same time, intensities decreased significantly, especially at high-energy. As a consequence, the estimated spectrum is better reproduced by a truncated power-law (E-R function), i.e. without a transition to a high-energy spectral index, so no value of $\gamma_{b}$ during this phase is reported in Figure \ref{fig:cParTime_Sept06}.

A complex temporal evolution characterized the initial phase of the 10 September event. During the first three time bins, the spectrum was relatively hard with $\gamma_{a}$ almost constant ($\sim$0.6) and $\gamma_{b}$ very slowly increasing. At the same time, two peaks were observed in the intensity profiles of the HEPADs; a minimum of the rollover energy $E_{r}$ and a maximum of the break energy $E_{0}$ were found in the interval between the peaks (20--23 UT). As discussed in section \ref{Near-Earth observations}, there may be alternative interpretations of this feature. In particular, the event commenced in the recovery phase of the FD, while the Earth was in a ICME region, and the second peak occurred after the arrival of a HSS following the trailing edge of the ICME.
The SEP event lasted for several days, with a monotonic increase of a $\gamma_{b}$ and, hence, a gradual softening of the spectrum, as the intensities of the higher energy particles accelerated earlier and closer to the Sun decline. The break energy remained relatively stable, within uncertainties, around a value of $\sim$20 MeV. After 13 September the rollover energy was probably higher than the maximum explored energy, and the spectra were better reproduced by the Band function. A significant suppression of intensities was registered as a consequence of the arrival of a HSS on 14 September which terminated the event at low energies and caused an abrupt increase of $\gamma_{b}$ from 5 to 7. Starting on 16 September the derived spectra between 2 and a few tens of MeV can be described by a simple power-law gradually approaching the background intensities, so results are not reported in Figure \ref{fig:cParTime}.

\begin{figure}[!t]
\centering
\begin{tabular}{c}
\includegraphics[width=0.73\textwidth]{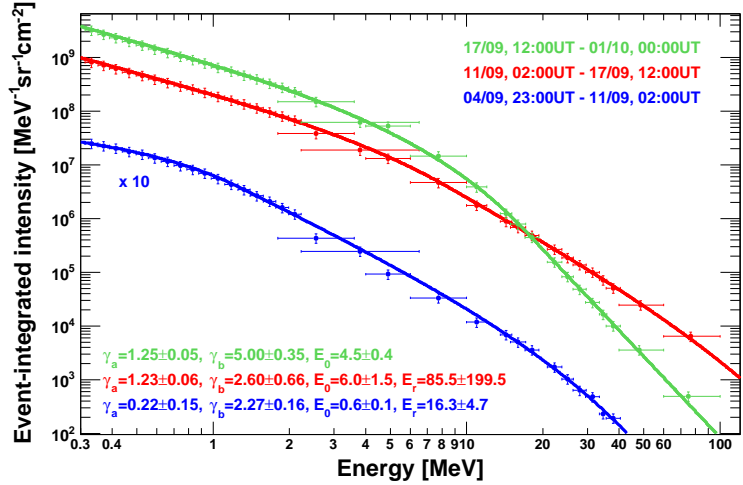}\\
\end{tabular}
\caption{Time-integrated energy spectra of the 4, 10 and 17 September 2017 SEP events (blue, red and green points respectively) measured by STEREO-A. 
The vertical error bars account for statistical and systematic uncertainties.
The horizontal error bars show the nominal range of each energy channel. The curves represent the fits based on the Band (for the 17 September event) and the combined (for the 4 and 10 September events) functions. The integration intervals, along with the fit parameters and associated uncertainties are also reported with the same color code.}
\label{fig:cStereoFluence}
\end{figure}

\subsection{Comparison with STEREO-A spectra}
Figure \ref{fig:cStereoFluence} displays the time-integrated energy spectra of the 4, 10 and 17 September events measured by STEREO-A (see Section \ref{Stereo observations}), denoted by blue, red and green points respectively. The spectra extend over the full energy range (300 keV -- 100 MeV) covered by the SEPT, LET and HET instruments. The curves represent the fits based on the Band (for 17 September event) and the combined (for the 4 and 10 September events) functions. The integration intervals, along with the fit parameters and associated uncertainties are also reported with the same color code.
The spectrum derived for the 4 September event is much less intense, and was multiplied by 10 to improve the comparison.
Albeit data points are limited to 40 MeV, it exhibits a break at very low energies ($E_{0}$$\approx$0.5 MeV) along with a rollover at higher energies ($E_{r}$$\approx$16 MeV).
In contrast, the spectra of the other two events extend above 60 MeV. While the high-energy data of the 10 September event spectrum suggest a rollover corresponding to $E_{r}$$\approx$79 MeV, although affected by very large uncertainties due to the limited number of points,
the spectral shape of the 17 September event is significantly softer above the break energy ($\gamma_{b}$$\approx$5); consequently, no rollover can be identified and the data are well reproduced by the Band function.
However, it should be noted that measured intensities include a contribution from the previous event that is apparently larger at lower energies. In addition, 
a component of low-energy particles is associated with the interplanetary shock arriving on 19 September (see Section \ref{Stereo observations}). 
Finally, the spectrum is influenced by the FD caused by the subsequent ICME, whose effects are not accounted for in the background subtraction,
as described in Section \ref{SEP spectral analysis}.

\begin{figure}[!t]
\centering
\includegraphics[width=0.73\textwidth]{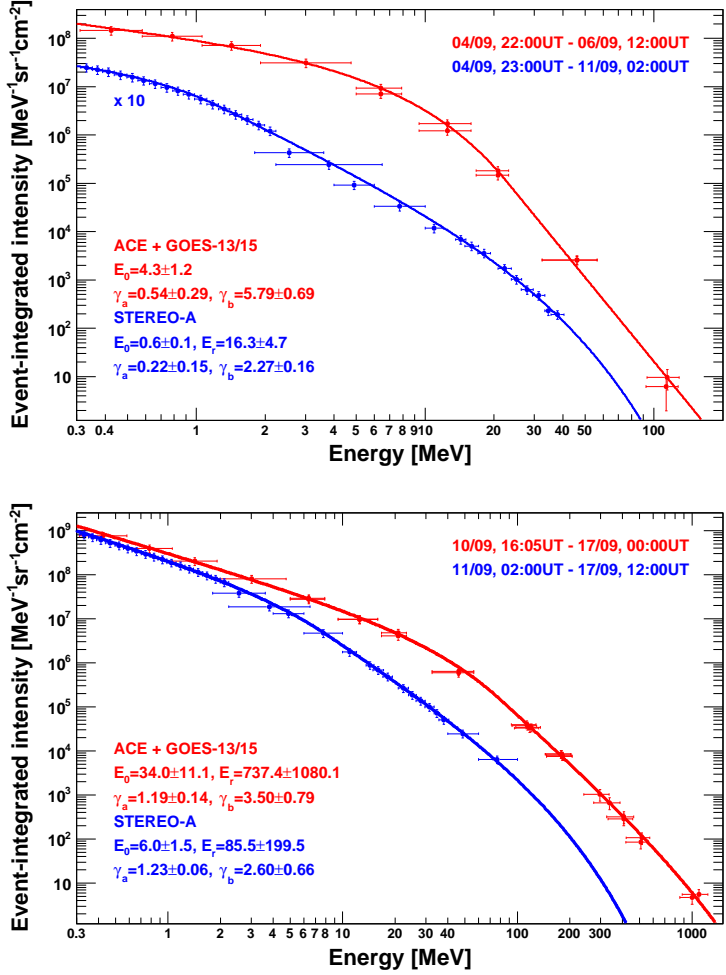}
\caption{Comparison between the time-integrated energy spectra measured by ACE and GOES-13/15 (red), and by STEREO-A (blue), during the 4 and 10 September 2017 SEP events (top and bottom panel, respectively). 
The vertical error bars account for statistical and systematic uncertainties; the horizontal error bars denote the channel nominal/effective energy ranges.
The curves are the fits based on the combined functional form (Equation \ref{eq:BANDER}) and, in case of the 4 September event spectrum measured by ACE and GOES, on the Band function (Equation \ref{eq:BAND}).
The integration intervals, along with the fit parameters and associated uncertainties are also displayed with the same color code. The STEREO-A spectrum derived for the 4 September was multiplied by 10 to improve the comparison.}
\label{fig:cFluence_fitcomp_stereo}
\end{figure}

Figure \ref{fig:cFluence_fitcomp_stereo} shows the comparison between the time-integrated energy spectra measured by ACE and GOES-13/15 (red), and by STEREO-A (blue), during the 4 and 10 September SEP events (top and bottom panel, respectively). 
The curves are the fits based on combined functional form and, for the 4 September event spectrum measured by ACE and GOES, on the Band function. In case of STEREO-A, the fit are extrapolated beyond the 100 MeV limit of the observations. 
The integration intervals, along with the fit parameters and associated uncertainties are also displayed with the same color code. 
Overall, the spectra differ in both magnitude and shape. In particular, the SEP events are larger near the Earth and their spectra extend to higher energies. 
Discrepancies are emphasized during the 4 September event, with a $\sim$100 factor for the time-integrated intensities at 1 MeV, while are less evident during the 10 September event. Such differences can be mostly attributed to the different magnetic connection of the spacecraft:
for both events, ACE and GOES footpoints were best connected to the solar event, detecting higher particle intensities and harder spectra (see, e.g., \citet{ref:HU2017}). On the other hand, STEREO-A was connected to the back side of the Sun (see Figure \ref{fig:spacecraft_positions}) and, as suggested by SEPMOD simulations \citep{ref:LUHMANN2018}, for the 10 September event it may have predominantly detected particles streaming from the distant shock beyond 1 AU (see Section \ref{Stereo observations}). 
STE\-REO observations demonstrate that this event was very broad in longitude even at high energies.
A major role was likely played by transport effects such as cross-field diffusion and IMF corotation, possibly in combination with widespread particle sources associated with a CME-driven shock accelerating and injecting particles onto an extended region of the heliosphere (see, e.g., \citet{ref:RICHARDSON2014,ref:LARIO2017} and references therein).
Additional factors should be considered when comparing the two sets of measurements, including the effects of SW structures.
In particular, near-Earth observations of the 10 September event were influenced by the interplanetary counterpart of the 6 September CME and the subsequent HSS (see Section \ref{Near-Earth observations}). 
We also note that measured SEP time-integrated spectra include a component from previous events and that the used integration intervals are limited by the onset of the subsequent events, e.g. the commencement of the 17 September event in case of STEREO-A.

\subsection{Comparison with the 17 May 2012 GLE event}
\begin{figure}[!t]
\centering
\begin{tabular}{c}
\includegraphics[width=0.73\textwidth]{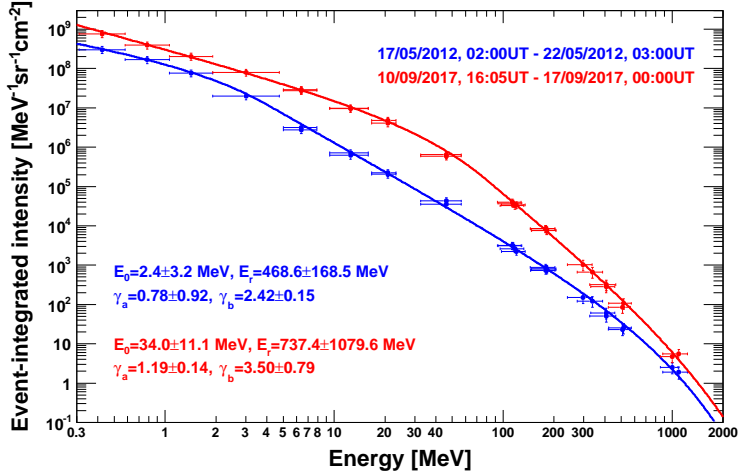} \\
\end{tabular}
\caption{Time-integrated energy spectra of the 17 May 2012 (blue) and the 10 September 2017 (red) SEP events measured by ACE and GOES. The vertical error bars account for statistical and systematic uncertainties. 
The horizontal error bars denote the channel nominal/effective energy ranges.
The curves are the fits based on the combined functional form (Equation \ref{eq:BANDER}).
The integration intervals, along with the fit parameters and associated uncertainties are also reported with the same color code.}
\label{ref:FitComp2012may17}
\end{figure}

Figure \ref{ref:FitComp2012may17} compares the time-integrated energy spectrum of the 10 September 2017 event (red) with that of the 17 May 2012 event (blue), associated with the previous GLE (n.71) of the solar cycle 24 \citep{ref:MAY17PAPER}. Both spectral fits, based on the combined functional form, rely on ACE and GOES observations according to the procedure described in Section \ref{SEP spectral analysis}. The integration intervals along with derived fit parameters and related uncertainties are also shown with the same color code. 
While the discrepancy in the absolute intensities reflects the much shorter duration of the 17 May 2012 event, the two spectral shapes are quite different, with the 10 September 2017 event exhibiting a softer spectrum above several tens of MeV, with higher break and rollover energies. 
This is consistent with PAMELA measurements \citep{ref:BRUNO2018}, showing that higher energy rollovers tend to be associated with larger spectral indices.
Based on a simple power-law fit of the data points above the transition energies (78.4 MeV and 3.9 MeV, respectively), a spectral index value of 4.05$\pm$0.03 and 2.97$\pm$0.20 is obtained for the 10 September 2017 and the 17 May 2012 events, respectively.

In general, several concomitant factors potentially contribute to the differences in the observed spectral shapes, such as the parent flare and CME parameters, the shock morphology and evolution, the ambient conditions, the magnetic connection to Earth and the interplanetary transport. The 17 May 2012 GLE event was peculiar because of the moderately strong source: an M5.6 flare linked to a 1582 km s$^{-1}$ linear speed CME in the CDAW catalog.
Such values are significantly lower compared with those associated with the 10 September 2017 event (X8.2 and 3163 km s$^{-1}$). However, the former event originated in a region characterized by a better longitudinal connectivity to Earth (N11W76) than the latter event (S08W88), and the 10 September 2017 flare reached peak intensity when the involved AR had just rotated over the western solar limb.
In addition, \citet{ref:GOPALSWAMY2018} proposed that the non-radial motion of the CME along with the favorable B0 angle (the inclination of the solar equator to the ecliptic) rendered the shock nose latitudinally well connected to Earth in case of the 17 May 2012 event, while the opposite situation occurred during the 10 September 2017 event.
Consequently, it can be speculated that the protons detected near the Earth at highest energies were accelerated mostly at the eastern flank of the shock, where acceleration is less efficient and 
the SEP maximum energy is lower \citep{ref:HU2017}, resulting in a softer spectrum with respect to better connected events such as the 17 May 2012 event. 

However, the prevailing interplanetary conditions may significantly complicate such arguments based on simple assumptions for the connectivity.
For instance, according to \citet{ref:ROUILLARD2016} the magnetic connectivity between the 17 May 2012 solar event and the near-Earth environment was established via a MC that erupted from the same AR a few days before.
Similarly, the 10 September 2017 event commenced while the Earth was in a ICME region, during the recovery phase of a FD causing a depression in observed intensities. Since the applied GCR background correction does not account for such effects being based on the average intensities registered prior to the three SEP events (see Section \ref{SEP spectral analysis}),
derived high-energy SEP intensities are somewhat underestimated.
In addition, the double-peak feature exhibited by the temporal profiles of high-energy intensities may be related to the influence of SW structures on particle transport.
Finally, measured time-integrated intensities include a low-energy contribution from the previous SEP event on September 6.
Consequently, the ``true'' SEP spectrum is supposed to be harder.

%
%

\section{Summary and conclusions}\label{Summary and conclusions}
Despite the near solar minimum conditions, an exceptional interval of solar activity occurred between 4 --10 September 2017 during the late decay phase of solar cycle 24 that involved the complex AR NOAA 12673 located in the western solar hemisphere. A large number of bright eruptions were observed, including four associated with X-class flares. The X9.3 flare on 6 September and the X8.2 flare on 10 September are currently the two strongest soft X-ray flares of solar cycle 24. Both were linked to fast CMEs, giving rise to SEP events measured by near-Earth spacecraft. In particular, the western limb event on 10 September triggered a GLE recorded by several NM stations, the second GLE (no.72) of the solar cycle. A further, smaller SEP event, detected late on 4 September, originated from the M5.5 flare and the related CME that erupted on the same day.

In this work we analyzed the space-based proton measurements by ACE and GOES-13/15 to study the time integrated spectra and spectral evolution of in a wide energy range ($\ge$300 keV).  
The spectra show a low-energy spectral break at few/tens of MeV, that is often attributed to the limits of diffusive shock acceleration, though interplanetary transport may also introduce such features in SEP spectra.
In addition, the 10 September 2017 event spectrum, extending up to $\sim$1 GeV, exhibits a high-energy rollover similar to that reported in the recent SEP observations of the PAMELA experiment, that may be ascribed to the limited extension and lifetime of the shock in the scenario of diffusive shock acceleration.
However, for the September 2017 period, the study of SEP features, including the interpretation of spectra shapes, is significantly complicated by a series of overlapping events and interplanetary structures (local shocks, ICMEs and HSSs), that influenced SEP intensities and hence the spectra. Consequently, it is not realistic to account for the spectral features only in terms of particle acceleration. 
In addition, a double peak in the high-energy proton intensity profile during the 10 September may have originated from a change in the connection conditions as the Earth moved from an ICME into a HSS; available radio burst data disfavor the alternative interpretation of a second particle injection.

Near-Earth SEP observations for these events have been compared with those reported by STEREO-A. 
Furthermore, we compared the spectrum for the 2017 September 10 event with that obtained for the 2012 May 17 event, associated with the previous GLE in cycle 24.
Differences in the spectra and their temporal evolution can be mostly attributed to the different magnetic connection of the spacecraft with respect to the shocks accelerating particles, 
but local interplanetary structures such as shocks, ICMEs and HSSs also have a relevant impact. STEREO data demonstrate that the 10 September 2017 event was very broad even at high energies, suggesting significant transport effects such as cross-field diffusion and IMF corotation in combination with the extended SEP source provided by the CME-driven shock.

%
%

\acknowledgments
The authors thank the ACE, GOES and STEREO teams for making their data publicly available.
A.~B. acknowledges support by an appointment to the NASA postdoctoral program at the NASA Goddard Space Flight Center administered by Universities Space Research Association under contract with NASA. 
I.~G.~R. acknowledges support from NASA Living With a Star grant NNG06EO90A.
Supporting data are included as six tables in an SI file.

%
%


\end{document}